\documentclass{article}%
\pdfoutput=1 
\usepackage[margin=2 cm]{geometry}
\usepackage{graphicx}
\usepackage{upgreek}
\usepackage{multicol,multirow}
\usepackage{amsmath,amssymb,amsfonts}
\usepackage{mathrsfs}
\usepackage{amsthm}
\usepackage[figuresright]{rotating}
\usepackage{appendix}
\usepackage[authoryear]{natbib}
\usepackage{ifpdf}
\usepackage[T1]{fontenc}
\usepackage{newtxtext}
\usepackage{newtxmath}
\usepackage{textcomp}
\usepackage{xcolor}
\usepackage{subcaption}
\usepackage{epstopdf}
\usepackage{psfrag}
\usepackage{pgfplots}
\usepackage{siunitx}
\usepackage{import}
\usepackage{float}
\usepackage{gensymb}
\pgfplotsset{compat=1.16}
\graphicspath{{exp_figs}}
\usepackage{authblk}

\usepackage[colorlinks,allcolors=blue]{hyperref}
\definecolor{jourcolor}{cmyk}{1,0.57,0.01,0.38}
\hypersetup{
    colorlinks,%
    citecolor=jourcolor,%
    filecolor=jourcolor,%
    linkcolor=jourcolor,%
    urlcolor=jourcolor
}

\newcommand{\ud}{\mathrm{d}}

\begin{document}

\title{Uniformly distributed floor sources of buoyancy can give rise to significant spatial inhomogeneities within rooms}

\date{}
\author[1,2]{Carolanne V.\ M.\ Vouriot}
\author[1]{Thomas D.\ Higton}
\author[2]{P.\ F.\ Linden}
\author[1]{Graham O.\ Hughes}
\author[1]{Maarten van Reeuwijk}
\author[1,*]{Henry C.\ Burridge}
\affil[1]{Department of Civil and Environmental Engineering, Skempton Building, South Kensington Campus, Imperial College London, London SW7 2BX, UK.}
\affil[2]{Department of Applied Mathematics and Theoretical Physics, Centre for Mathematical Sciences, University of Cambridge, Wilberforce Rd, Cambridge CB3 0WA, UK.}
\affil[*]{Corresponding author: h.burridge@imperial.ac.uk}

\maketitle

\begin{abstract}

Displacement ventilation, where cool external air enters a room through low-level vents and warmer air leaves through high-level vents, is characterised by vertical gradients in pressure arising from the warmer indoor temperatures. Models usually assume that horizontal variations of temperature difference are small in comparison and are, therefore, unimportant. Small-scale laboratory experiments and computational fluid dynamics were used to examine these flows, driven by a uniformly heated floor. {These experiments and simulations show that the horizontal variations of temperature difference can be neglected for predictions of the bulk ventilation rate; however, they also evidence that these horizontal variations can be significant and play a critical role in establishing the pattern of flow within the room --- this renders the horizontal position of the low- and high-level vents (relative to one another) important.} We consider two cases: single-ended (where inlet and outlet are at the same end of the room) and opposite-ended. In both cases the ventilation flow rate is the same. However, in the opposite-ended case a dead zone is established in the upper part of the room which results in significant horizontal variations. We consider the formation of this dead zone by examining the streamline patterns and the age of air within the room. We discuss the implications for occupant exposure to pollutants and airborne disease.

\textbf{\mathversion{bold}Impact Statement} Exposure to indoor air pollution and airborne diseases are major factors in human health and well being. Guidance on appropriate ventilation rates is typically based on bulk ventilation rates, either in terms of the amount supplied per individual or as air exchange rates for a space. Such bulk measures assume homogeneous conditions within a space while, in practice, there are often significant spatial variations in properties. This paper shows that in displacement ventilation, where it is commonly assumed that horizontal variations are negligible, in fact simply altering the horizontal position of an outlet vent can lead to large variations in indoor air quality. Consequently, exposures based on average values can be misleading. This finding also has important implications for the location of sensors to measure the conditions of indoor environments, something which is becoming increasingly commonplace. 


\end{abstract}
\section{Introduction} \label{sec:intro}

\begin{figure}[]
	\centering
	\def\svgwidth{0.6\textwidth}
\begingroup%
  \makeatletter%
  \providecommand\color[2][]{%
    \errmessage{(Inkscape) Color is used for the text in Inkscape, but the package 'color.sty' is not loaded}%
    \renewcommand\color[2][]{}%
  }%
  \providecommand\transparent[1]{%
    \errmessage{(Inkscape) Transparency is used (non-zero) for the text in Inkscape, but the package 'transparent.sty' is not loaded}%
    \renewcommand\transparent[1]{}%
  }%
  \providecommand\rotatebox[2]{#2}%
  \newcommand*\fsize{\dimexpr\f@size pt\relax}%
  \newcommand*\lineheight[1]{\fontsize{\fsize}{#1\fsize}\selectfont}%
  \ifx\svgwidth\undefined%
    \setlength{\unitlength}{342.26975328bp}%
    \ifx\svgscale\undefined%
      \relax%
    \else%
      \setlength{\unitlength}{\unitlength * \real{\svgscale}}%
    \fi%
  \else%
    \setlength{\unitlength}{\svgwidth}%
  \fi%
  \global\let\svgwidth\undefined%
  \global\let\svgscale\undefined%
  \makeatother%
  \begin{picture}(1,0.53509965)%
    \lineheight{1}%
    \setlength\tabcolsep{0pt}%
    \put(0,0){\includegraphics[width=\unitlength,page=1]{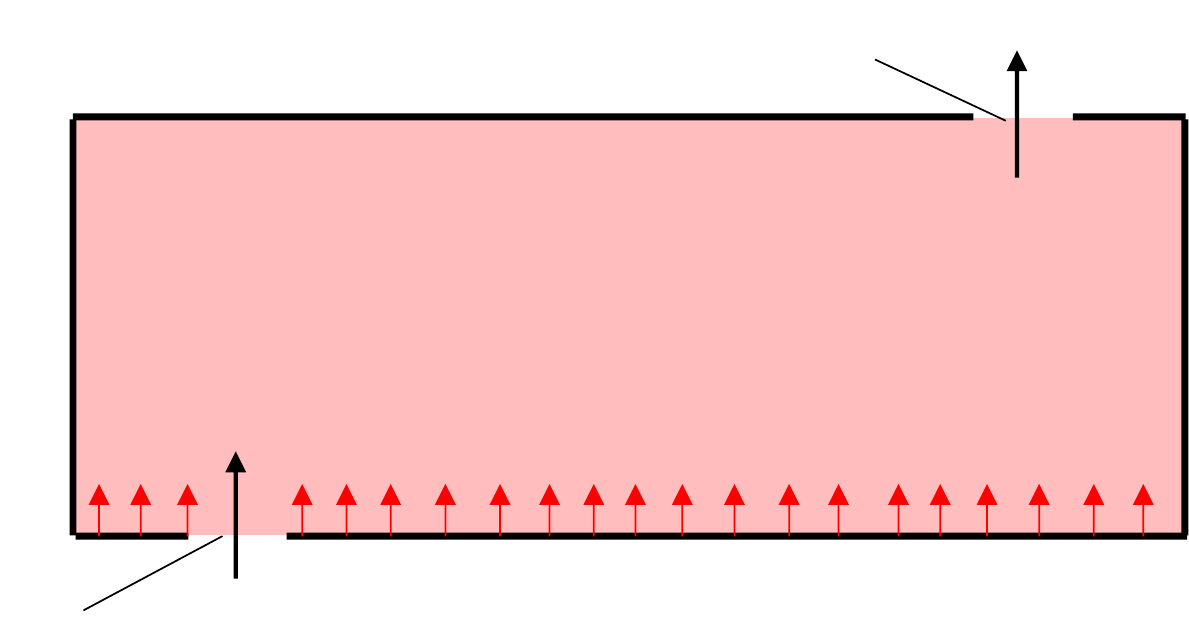}}%
    \put(0.86586602,0.47391533){\makebox(0,0)[lt]{\lineheight{1.25}\smash{\begin{tabular}[t]{l}$Q$\end{tabular}}}}%
    \put(0.68880872,0.49599093){\makebox(0,0)[lt]{\lineheight{1.25}\smash{\begin{tabular}[t]{l}$A_h$\end{tabular}}}}%
    \put(0.20562914,0.02741461){\makebox(0,0)[lt]{\lineheight{1.25}\smash{\begin{tabular}[t]{l}$Q$\end{tabular}}}}%
    \put(0.01944752,0.00937277){\makebox(0,0)[lt]{\lineheight{1.25}\smash{\begin{tabular}[t]{l}$A_l$\end{tabular}}}}%
    \put(0.49684214,0.26159863){\makebox(0,0)[lt]{\lineheight{1.25}\smash{\begin{tabular}[t]{l}$\overline{\Delta\mathcal{T}_m} = \textrm{uniform}$\end{tabular}}}}%
    \put(0.49934387,0.13662459){\makebox(0,0)[lt]{\lineheight{1.25}\smash{\begin{tabular}[t]{l}$F$\end{tabular}}}}%
  \end{picture}%
\endgroup%

	\caption{Schematic of the traditional well-mixed model for a room ventilated, via low- and high-level vents, by a flow driven by distributed floor source of heat (buoyancy). Our research shows this view to only be appropriate for predicting bulk ventilation rates.}
	\label{fig:schem}
\end{figure}
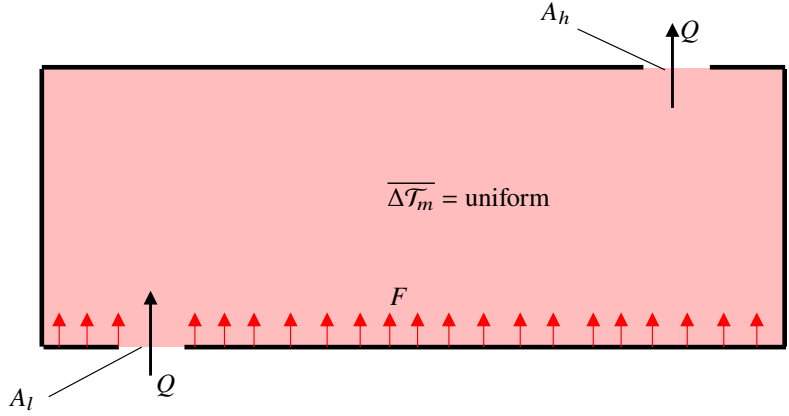

Where feasible, displacement ventilation strategies, i.e. the low-level introduction and high-level extraction of air within building spaces, offer the potential for reduced energy usage \citep{Linden99,Wachenfeldt07} and improved indoor air quality \citep{Sandberg81,Bhagat20}.
The physics of displacement ventilation strategies are long studied \citep[e.g.][]{Linden90,Gladstone01} with the standard assumptions that spaces are horizontally uniform; this assumption implies that the horizontal location of ventilation openings is irrelevant. We seek to highlight that this is not always the case, even in the limit where the bulk vertical ventilation flow within a space is naturally driven by buoyancy forces (arising from ubiquitous temperature differences) without any enhancement by external winds. In such cases predictions of the ventilation flow, and the indoor conditions, are typically made by assuming the space to be either well-mixed or stratified in layers (two-layers in the simplest case). Which arises, is broadly taken to depend on whether the low-level heat (buoyancy) sources can be considered to be (horizontally) distributed or localised \citep{Linden99}, with localised heat sources taken to generate coherent vertical flows, i.e. plumes. In this latter case, the bulk vertical ventilation flow is carried across interface(s) by these plumes; knowledge of the vertical evolution of the volume flux within the plumes can be combined with the location of the interface(s) to provide estimates of the ventilation flow rate. Our focus is the former case, i.e. displacement ventilation driven by distributed heat sources, and we challenge the traditional view (see figure \ref{fig:schem}) of the room conditions that arise.

It is well known that horizontal fluid motion can play a significant role in determining the transport in confined regions of heated fluid \citep{Hughes08}, yet such knowledge is rarely considered when designing indoor spaces. The displacement ventilation flow and layered stratification generated in the presence of the heat input from a single person within a room (of albeit limited size) have been investigated via high resolution direct numerical simulations in the case that the inflow is mechanically forced \citep{Yang22}. With this freedom to force the horizontal inflow, \citet{Yang22} showed that by altering the horizontal position of the upper outlet, from one side of the room to the other with respect to the inlet, the height of the interface was altered, {and} that this height {is} critical {to} predict where exposures within a space might be high. Studies of naturally driven  displacement ventilation have also shown the ability of external wind to force the flow to transition, from one which maintains a layered stratification, to one in which horizontal motions mix the space when the wind is sufficiently strong. For example, \citet{Hunt99} define a Froude number condition to describe the relative forcing of wind to buoyancy; above a threshold Froude number, one can expect layered stratifications to become mixed. We assert, that in cases when layered stratifications are disrupted, while well-mixed models might still be capable of predicting bulk flow rates with suitable accuracy, predicting exposures based on horizontal uniformity may be risky. This has already been evidenced by \citet{Yang22} for ventilation flows that can be forced arbitrarily hard. However, here we show our assertion to have wider validity by investigating displacement flows `naturally' driven only by distributed floor sources of buoyancy. {Although} such sources are widely expected to result in conditions reasonably approximated by the well-mixed assumption, e.g. figure \ref{fig:schem}, we show that this is not the case.

In the current study, we deploy both experiments and simulations, of ventilation flows through low- and high-level vents driven by uniformly distributed floor sources of buoyancy (equivalent to the heat loads in operational rooms during typical heating season), to examine the distribution of buoyancy (temperature), and associated flow fields within idealised empty rooms. We examine the impact of changing room aspect ratio and changing the relative horizontal position{s} of the low-level inlet and high-level outlet vents, so that in some configurations the flow exits the room in the opposite corner to the inlet and, in other configurations, the flow exits the room at the same end of the room as the inlet. We choose to describe these as 1) opposite-ended configurations, and 2) single-ended configurations, {respectively}. We note that within our terminology we attempt to draw a distinction from the terms `cross-flow ventilation' (sometimes simply `cross ventilation') and `single-sided ventilation', which are taken to refer (implicitly in the case of `cross-flow', and explicitly for `single-sided') to ventilation flows through openings in walls \citep[see][]{CIBSEAM10}, and are sometimes taken to refer to wind dominated ventilation. However, we note that our terminology for the configuration is deliberately not explicit in distinguishing between vertically (i.e. in walls) or horizontally (i.e. in ceilings or floors) aligned openings as, with some considered thought, our findings can be applied to both.

The remainder of the paper is structured as follows. In \S \ref{sec:methods} we describe the fundamental parameters which determine the scale of the bulk ventilating flow. Details of the laboratory (\S \ref{sec:exp}) and numerical (\S \ref{sec:num}) experiments are presented before moving on to our results, \S \ref{sec:res}. We present results for the bulk flow rates attained, and then focus on {the }buoyancy field within the room from our experiments in \S \ref{sec:exp_obs}. These are then compared and contrasted to those of our simulations, first examining the effect of relative horizontal position of the inlet and outlet vents {in} \S \ref{sec:rel}, before investigating the fluids mechanics which underpins such findings (\S \ref{sec:tangle}), and then considering the implications for occupant experience within \S\ref{sec:impli}. Finally, we draw conclusions (\S\ref{sec:conc}).

\section{Methodology} \label{sec:methods}

This study was inspired by observations made during an experimental campaign, which were then supplemented by a set of numerical simulations, of the natural ventilation flow in a room with low- and high-level vents separated by a vertical distance, $H$. Within both our experiments and simulations the flow is driven by a buoyancy source, uniformly distributed over the floor, within a room connected to the external environment by a low-level inlet and a high-level outlet --- such set-ups are typically expected to generate relatively well-mixed conditions within the room \citep{Linden99}.

Bulk ventilation flow rates $Q$ were created within the room due to the stack pressure, arising due to the buoyancy, over the height {$H$}. In the case that vents are vertically aligned (e.g. a door-like ventilation opening) we measure $H$ from the half-height of the opening. {These flows} were generated {by an} integral source buoyancy flux $F$ (or equivalently an integral heat flux $W$; with $W = \rho \, c_p  \, F / (\beta \, g)$, where $g$ is the gravitational acceleration, and $\rho$, $c_p$, and $\beta = 1/T_{ref} \approx 1/300\,\textrm{K}$ are the fluid density, specific heat capacity, and thermal expansion coefficient, respectively. In all cases the buoyancy flux was input evenly over the {whole} floor. The flow is geometrically constrained by the low- and high-level vents of areas $A_l$ and $A_h$, respectively, providing resistance to the bulk flow via the effective area
\begin{equation}
    A^* = \frac{\sqrt{2} \; C_l \, A_l \, C_h \, A_h}{\sqrt{(C_l \, A_l)^2 + (C_h \, A_h)^2}} \approx \frac{\sqrt{2} \, C_d \, A_l \, A_h}{\sqrt{A_l^2 + A_h^2}} \, ,
\end{equation}
where $C_l$ and $C_h$ are the loss coefficients at the low- and high-level vents respectively, and the right-hand side becomes an equality when ${C_d \equiv}  C_l = C_h$.

For the case of a well-mixed room the uniform buoyancy is $F/Q$, and in the presence of low- and high-level vents, \citet{Linden99} provides the expected volume flux as 
\begin{equation}
    Q_{M} = \left( A^{*2} \, F \, H \right)^{1/3} \, . \label{eq:Qm}
\end{equation}
To aid comparison to the well-mixed state, we present our results via the dimensionless volume flux
\begin{equation}
    \hat{Q} = \frac{Q}{Q_{M}} \, ,
\end{equation}
and dimensionless time, scaled by the ventilation air change rate, namely
\begin{equation}
    \hat{t} = \frac{t}{V/Q}\,,
\end{equation}
where $V$ is the volume of the room.

Since our interest is to investigate in-room quantities we present measurements of the buoyancy $b=b(x,y,z,t) = [\rho_a - \rho(x,y,z,t)]g/\rho_a = \beta [ T(x,y,z,t) - T_a] \, g$, where $\rho_a$ and $T_a$ are the (uniform) density and temperature of the fluid within the ambient environment, respectively. {We note that temperature differences within the room are typically at least two orders of magnitude smaller than absolute temperatures, and hence the Boussinesq approximation is valid.} We use an overbar to denote quantities averaged over a time interval $t_{i}$, which is suitable to obtain good statistical estimates of the steady mean, e.g. $\overline{b}$. In order to present variations, within the room, about the `well-mixed' state, and because temperature is the active scalar of interest, we present the scaled buoyancy
\begin{equation}
    {\overline{\Delta \mathcal{T}}(x,y,z)} = \frac{\overline{b}(x,y,z)}{F/Q} = \frac{\rho \, c_p  \,[\overline{T}(x,y,z) - T_a]}{W/Q} \, ,
\end{equation}
so that $\Delta \mathcal{T}=1$ for a fully-mixed room. Finally, we utilise angle brackets, $\langle \boldsymbol{\cdot} \rangle$, for spatial averages. {For example, we describe data averaged over the (spanwise) width of the box as `width-averaged', denoting this data $\langle \overline{\Delta \mathcal{T}} \rangle_{y} = \overline{\Delta \mathcal{T}}(x,z)$.} Similarly, averages over horizontal cross-sections are denoted $\langle \overline{\Delta \mathcal{T}} \rangle_{x,y}\, = \overline{\Delta \mathcal{T}}(z)$, and averages over the room are denoted $\langle \overline{\Delta \mathcal{T}} \rangle_{V}$, with $\langle \overline{\Delta \mathcal{T}} \rangle_{V}=1$ for a {fully}-mixed room. {We note that the buoyancy data gathered via experiments (see \ref{sec:exp}) are effectively, integrated as light rays traverse the width of the box (i.e. in the direction of the coordinate we denote `$y$') --- a process which we describe as producing `width-averaged' data. In the numerical simulations, the width-averaged results are obtained by taking an average of 54 vertical planes across the room.}

\subsection{Laboratory experiments} \label{sec:exp}

Experiments were conducted using the salt-bath technique, where room-scale ventilation can be studied using small-scale models submerged in water \citep{Linden99}. The use of water as the fluid medium ensures that dynamic similarity, in both Reynolds and Rayleigh numbers, is broadly achieved at a room:model scale of around 10:1, and using salt (rather than heat) as the buoyancy scalar, as was realised in our experiments. Modelling the buoyancy, arising in buildings due to temperature differences in air, via saline differences in water also simplifies flow visualisation and enabled the observations included in this study. We choose to describe the experiments in the orientation of an actual room, such that we describe the fluid as being affected by a floor-level buoyancy source, rising up, and exiting via the high-level vent.

{A cuboid}, as a small scale model of a room, was connected to the ambient environment by two openings, a low-level inlet vent (aligned in the vertical plane much like a small doorway) and a high-level outlet vent, 
the position of the vents are illustrated in figure \ref{fig:exp_t_avg_buoyancy_fields} -- see table \ref{tab:params} for details.
{The high-level outlet was positioned in the ceiling approximately two thirds of the way along the room from the inlet, and hence the bulk flow must be both upwards and across the room.
A saline solution was supplied to the scale model at floor level across a porous plastic panel. The source solution was dyed, and the model back-lit, to enable the width-averaged two-dimensional buoyancy field $\langle b \rangle_y=b(x,z,t)$ to be measured using the dye-attenuation technique \citep{Cenedese97,Allgayer12}.}
Note that, experimental data directly adjacent to the walls are compromised due to parallax errors, camera viewing restrictions, and reflections from the walls and hence these regions are excluded from the results presented --- these regions are highlighted in figure \ref{fig:exp_t_avg_buoyancy_fields}. {Appendix \ref{app:exp} provides further details of the experimental set-up and procedures.}

{Our experiments, investigating the flow established by a buoyancy source (uniformly) distributed over the entire area, included the low-level vent being vertically aligned, akin to a doorway, but we only examined cases for which a unidirectional inflow was achieved at the low-level doorway.} We note that when presenting data from, and describing the flow field of, our experiments we have chosen to invert the set-up. This decision was taken to ensure a direction of the flow that is consistent with our numerical simulations and our primary application of interest, i.e. building ventilation; akin to this we choose to describe the confining experimental box as a `room'.

\subsection{Numerical simulations} \label{sec:num}

Numerical simulations were conducted with OpenFOAMv2106 using the transient \texttt{buoyant\-Pimple\-Foam} solver. Simulations were run using the Reynolds-averaged Navier-Stokes (RANS) $k-\omega$ SST turbulence model (\texttt{kOmega\-SST}). {RANS models only solve for the mean flow and model the remaining turbulent scales, as such they are not able to resolve all the flow features which might otherwise be resolved, for example, with Large Eddy Simulations. In the present work, however, the focus is on the overall flow pattern in a room set by large scale buoyancy variations, which RANS models are expected to capture. The relatively low cost of RANS was therefore exploited to consider different ventilation configuration and a range of vent areas. The $k-\omega$ SST model was specifically chosen for its ability to model flows near walls, shear flows and potential relaminarisation, all of which are needed to simulate ventilating flows. } 

A room domain, of size 10 \,m $\times$ 5.5\,m $\times$ 2.7\,m, was meshed and connected to larger exterior boxes through low- and high-level vents which both lay in horizontal plane{s} and with effective areas in the range $0.012 \leq A^*/H^2 \leq 0.049$, see table \ref{tab:params}. For all simulations discussed herein, the area of the high-level vent, $A_h$, was smaller than that of the low-level vent, $A_l$, falling in the range $0.25 \leq A_h/A_l \leq 0.5$ --- this avoided excessive momentum flux at the low-level vent (associated with unwanted cold drafts in rooms). For the example illustrated in \S \ref{sec:res}, $A_h/A_l = 0.5$. The larger exterior allowed the flow to be entirely driven by the buoyancy source by ensuring sufficient spatial separation between any prescribed boundary conditions and the flow within the room. The horizontal position of the high-level vent was varied from being at the far end of the room away from the low-level vent (much like the experimental configuration), to being at the same end of the room as the low-level vent; {as mentioned above} we describe these configurations as `opposite-ended' and `single-ended', respectively. A wintertime scenario was investigated where the external ambient temperature was set to $T_a=278$\,K (5\,\degree C) and the flow was driven by imposing a 6200\,W heat flux on the floor. An additional scalar transport equation was solved for a passive tracer representing the age of air $t_{AoA}$. Defined to be equal to 0 at the inlet, the age of air represents the time required to reach a certain point in the space and is often used to assess the efficiency of a ventilation system \citep{Sandberg81}. Herein we present the age of air scaled by the ventilation air change rate such that $\hat{t}_{AoA} = t_{AoA} V/Q$. {Appendix \ref{app:sim} provides further details of the numerical set-up.} 
 
\section{Results} \label{sec:res}

\renewcommand{\arraystretch}{1.35}
\begin{table}
    \centering
    \begin{tabular}{l|c|c}
         Parameter (range) & Experiments & Simulation \\
         \hline
         Room volume, $V\;[\textrm{m}^3]$ & 0.041 & 149 \\
         Room length, $L\;[\textrm{m}]$ & 0.45 & 10 \\
         Room width, $W\;[\textrm{m}]$ & 0.3 & 5.5 \\
         Vertical distance, $H\;[\textrm{m}]$ & 0.27 & 2.7\\
         Effective area, $A^*\;[\textrm{m}^2]$ & $0.8\times10^{-3}$ -- $1.8\times10^{-3}$ & 0.08 -- 0.33 \\
         Buoyancy flux, $F\;[\textrm{m}^4/\textrm{s}^{-3}]$ & $5.8\times 10^{-6} 
         $ & 0.17\\
         {Vent Reynolds number, $Re$ [-]} & { $3.36 \times 10^3 - 4.17 \times 10^3$} & { $3.95 \times 10^4 - 4.63 \times 10^4$}\\
         {Bulk Rayleigh number, $Ra$ [-]} & {$8.41 \times 10^{11} - 4.25 \times 10^{11}$} & {$8.09 \times 10^{10} - 3.40 \times 10^{10}$} \\
         Dimensionless volume flux, $\hat{Q}\;[-]$ & 0.95 -- 1.03 & 0.99 -- 1.07\\
    \end{tabular}
    \caption{Relevant parameters, and resulting flow rate, ranges spanned by our experiments and simulations. Note that in all of the above we take the discharge coefficient $C_d = 0.65$, Appendix \ref{app:sen} presents the sensitivity to this choice.}
    \label{tab:params}
\end{table}

Before examining the buoyancy field, and ultimately the flow patterns that influence it, we examine bulk ventilation flow rate attained. Table \ref{tab:params} shows details of the experimental and simulated {configurations considered.} {The breadth of the underlying physical parameter ranges explored is significant, and results in the vent Reynolds and bulk Rayleigh numbers each varying by at least an order of magnitude. For completeness, we define the vent Reynolds number as $Re = Q / (A^{*1/2} \, \nu)$, where $\nu$ is the kinematic viscosity of the fluid, and the bulk Rayleigh number $Ra = F \, H^{3}/(Q \, \kappa \, \nu) $, where $\kappa$ is the thermal or molecular diffusivity, as appropriate for the simulations or experiments, respectively. Despite the breadth of our investigation, the range of dimensionless volume fluxes attained, $\hat{Q}$, vary only slightly from unity (by always less than 7\%) --- this simply demonstrates that the realised ventilation flow rate is in good agreement with the classical theory based on the well-mixed assumption, and hence adheres to the scaling, namely (\ref{eq:Qm}). Moreover, the relatively narrow range of $\hat{Q}$ observed highlights that if one is only interested in predicting the bulk ventilation flow then one can conclude that the well-mixed model provides appropriate estimates --- at least to within uncertainty (much of which centres around the parameterisation of the losses at the vent via the coefficient $C_d$, see Appendix \ref{app:sen} for the sensitivity of $\hat{Q}$ on $C_d$).} 

In subsequent sections, we report detailed findings for a subset of the configurations reported in table \ref{tab:params}; for convenience of communication only, we use one representative example experiment and one simulation for consistent illustration. Crucially, examination of the scalar field shows that relatively well-mixed state, expected for indoor flows driven by distributed floor sources of buoyancy, was not observed, and this hitherto unreported observation forms the motivation of this study. We proceed to show that if one is required to predict the scalar field, for example, {to} predict occupant thermal {comfort} or exposure to pollutants, then the well-mixed model may be inadequate. 

\subsection{Experimental observations and measurements of the width-averaged buoyancy field} \label{sec:exp_obs}

\begin{figure}[] 
 	\centering
 	 \def\svgwidth{0.7\textwidth}
 	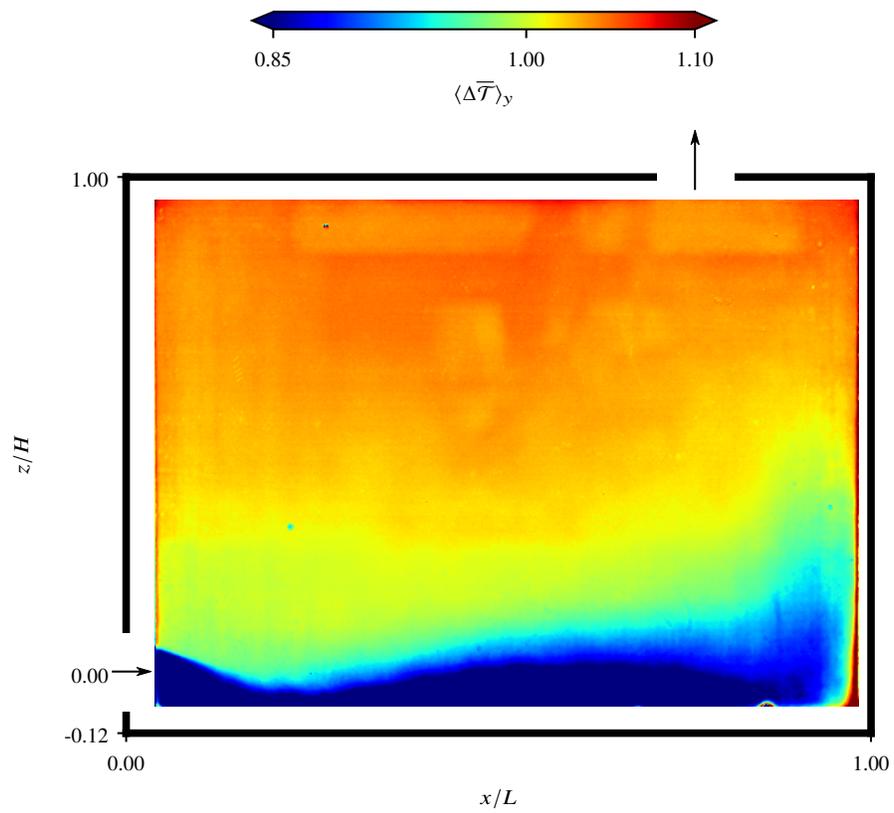
 	\caption{The width-averaged scaled buoyancy field $\langle \overline{\Delta \mathcal{T}}\rangle_y$ for the experiment (\S \ref{sec:exp}). The observations described based on the video images (\S \ref{sec:exp_obs}), that the flow field is strongly heterogeneous is evident in this image of the time-averaged buoyancy field.
 	}
 	\label{fig:exp_t_avg_buoyancy_fields}
\end{figure}

Figure \ref{fig:exp_t_avg_buoyancy_fields} shows the width-averaged (scaled) buoyancy field $\langle \overline{\Delta \mathcal{T}} \rangle_y$ from the, statistically steady-state, flow analysed. It is evident that buoyancy within the room is not {uniformly} mixed, e.g. there is a cooler region close to the floor, which rises at the far wall, and a warmer region in the upper part of the room. This results in spatial variation of buoyancy within the room spanning a range of around $20\%$ of the mean buoyancy. Moreover, the {horizontal} variation in buoyancy within the lower two-thirds ($0 \leq z/H \leq 2/3$) of the room is greatest. In many modern rooms ceilings are relatively low, and this region is therefore likely to be that experienced by occupants; thus, predicting these variations is a worthy research challenge.

We note that more significant inhomogeneities were observed in near wall regions which are compromised due to camera viewing restrictions, parallax, and reflections from the walls. The observations apparent in these images are more clearly evident in the video included as supplementary material. In this video, as with our other experiments, a statistically steady state had been attained before recording begun (the experiment had been already been running for over 9 air-change periods, i.e. a duration of $\hat{t} \approx 9.5$). However, unlike our other experiments {--- i.e. unlike those which provided our quantitative experimental results ---} the buoyant source fluid ejected was left undyed, {and} instead a sudden (low momentum) release of dyed neutrally buoyant fluid was made in the ambient environment just outside the inlet vent (by bursting a water filled balloon). This introduced a transient passive scalar (i.e. dye) concentration field within the statistically steady flow. The video illustrates the complexity of the flow field as relatively cool dyed fluid enters the room and travels along the floor as a gravity current that is warmed by the convection established above the floor buoyancy source. Fluid in the gravity current impacts the far wall, against which it rises (see the supplementary video and figure \ref{fig:exp_t_avg_buoyancy_fields}). A portion of this rising flow travels relatively directly to the outlet vent, while a portion of it remains in the upper part of the room {and a further} portion slumps to form an intruding gravity current above, and oppositely-directed to, the incoming current. {At the same time} large-scale convective motion is evident throughout the entire room.

The similarities of th{is} flow {to those driven by} `horizontal convection' \citep{Hughes08} are apparent (as discussed in \S\ref{sec:tangle} below). However, it is unclear the extent to which the vertical alignment of the low-level doorway, and the momentum associated with the horizontal inflow, might be critical in establishing the flow field that results in the horizontal convection observed. It had been established \citep{Yang22} that when the inflow is forced sufficiently hard horizontally, relative to the stratification generated by a localised buoyancy source, then the canonical flow patterns expected are disrupted. {To enable efficient examination of a wider range of conditions, including: a different room aspect ratio, eliminating the horizontal momentum of the incoming flow, and altering the relative horizontal positions of the low- and high-level vents, we now turn to the results from our investigation via numerical simulations.}

\subsection{Numerical results for different relative inlet to outlet vent positions} \label{sec:rel}

The horizontal momentum of the inflow was eliminated within {the} simulations by choosing to align {the} low-level vent horizontally within the `floor' of {the} room domain so that the incoming ventilation flow is purely vertical. Moreover, to test the sensitivity of our findings to the room aspect ratio we simulated a room of $L/H \approx 4.35$, \textit{cf.} $L/H \approx 1.50$ within the experiments.

\begin{figure}
\centering
     \begin{subfigure}[h]{\textwidth}
         \centering
         \includegraphics[width=\textwidth]{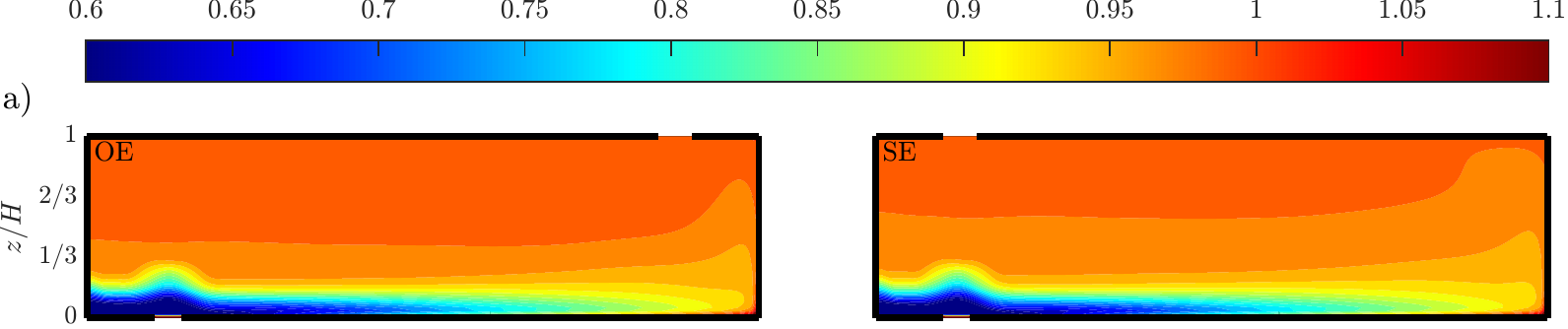}
     \end{subfigure}
     \hfill
    \vspace{11pt}
         
         \begin{subfigure}[h]{\textwidth}
         \centering
         \includegraphics[width=\textwidth]{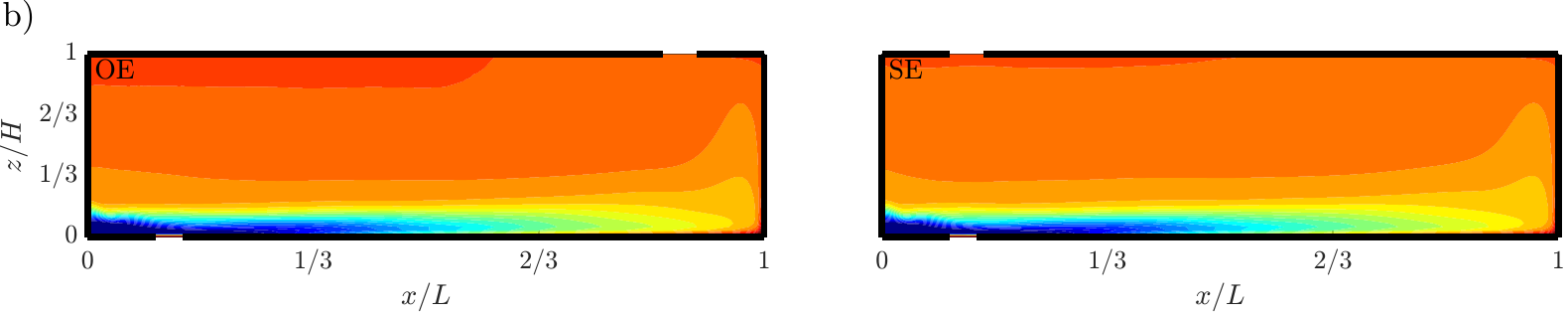}
     \end{subfigure}

        \caption{Scaled buoyancy fields from the numerical simulations in which a uniformly distributed heat source at floor level drives the flow through the (horizontally aligned) low- and high-level vents, located on the floor and ceiling of the room, respectively. The results presented are independent of the choice of parameters within the ranges shown in table \ref{tab:params} but for the simulation shown, $A_l= 0.030\,\textrm{m}^2$ and $A_h= 0.015\,\textrm{m}^2$, with $H=2.7\,\textrm{m}$, taking $C_d=0.65$ gives $A^*/H^2 = 0.012$. The top row presents, a) the {width}-averaged buoyancy field $\langle \overline{\Delta \mathcal{T}}\rangle_y$, \textit{cf.} figure \ref{fig:exp_t_avg_buoyancy_fields}, and the bottom row presents, b) the scaled buoyancy field at the central plane $\overline{\Delta\mathcal{T}}(y = W/2)$. In each {row}, the opposite-ended {(OE) }configuration is shown in the left column and the single-ended {(SE) }configuration in the right column. Note that thick black lines mark the confined edges of the room, gaps {show} the positioning of the low- and high-level, inlet and outlet, vents respectively. {Note that, as with the other figures included herein, data is only shown from within the room, see figure \ref{fig:Mesh} for an illustration of the room within the full computational mesh.}}
        \label{fig:CFD_buoyancyfield} \label{fig:CFD_buoyancyaveraged}
        \label{fig:CFD_buoyancycentralaveraged}
\end{figure}

Figure \ref{fig:CFD_buoyancyfield} shows the {width-averaged} buoyancy field obtained from our numerical simulations. The left-hand column plots results from the opposite-ended configuration (in which the low- and high-level vents are positioned in the region of opposite corners of the room) and the single-ended configuration {results are shown in the right-hand column} (in which the vents are both positioned near the left-hand end of the room, i.e. closer to $x/L=0$, but with the low-level inflow vent near the back of the room (being $y/W=1$) and high-level outflow near the front of the room (being $y/W=0$)). {The {flow} patterns of the simulations (figure \ref{fig:CFD_buoyancyaveraged}) are comparable to those of the experiments (figure \ref{fig:exp_t_avg_buoyancy_fields}), with significant spatial variations in the buoyancy field especially in the bottom third of both rooms, and are robust to the significant variation in aspect ratio between rooms (depicted to scale within these figures) and to the fact that in our simulations {the} incoming flow was vertical}. This {agreement }suggests {that }the horizontal momentum of the inflow in our experiments does not play a significant role in establishing the dynamics within the room, and these dynamics are robust to changes in aspect ratio (at least over the range 1.50--4.35, which spans many geometries observed in rooms). Moreover, differences between the experimental data and that of the numerical simulations are of a similar magnitude to the difference{s} that arise when the position of the upper vent in the simulations is moved from being nearer the right-hand wall furthest from the inlet, i.e. {the} opposite-ended configuration (as was the case in {the} experiments), to being nearer the left-hand wall, i.e. single-ended. For example, in {the} opposite-ended simulations, there is an area in the upper part of the room in which the buoyancy (temperature) is greater than any observed in {the} single-ended configuration -- this occurs predominantly at the inlet (left-hand) side of the room.

{F}igure \ref{fig:CFD_buoyancycentralaveraged} {shows} the buoyancy field on the central {vertical }plane ($y/W=0.5$), {which exhibits} broadly the same features as th{e} {width}-average, but with slightly greater variation between the two flow configurations and more evidence of convection across the plane. We return {to} the variations between different regions of the room in \S \ref{sec:impli}.

The variation between the simulated opposite-ended and single-ended configurations are quite marked given that the two simulations are expected to give rise to a well-mixed room \citep{Linden99}, and the configurations are notionally identical except for the horizontal location of the upper vent. We now consider the physics which give rise to these observations.

\subsection{Untangling the flow} \label{sec:tangle}

\begin{figure}
\centering
     \begin{subfigure}[h]{\textwidth}
         \centering
         \includegraphics[width=\textwidth]{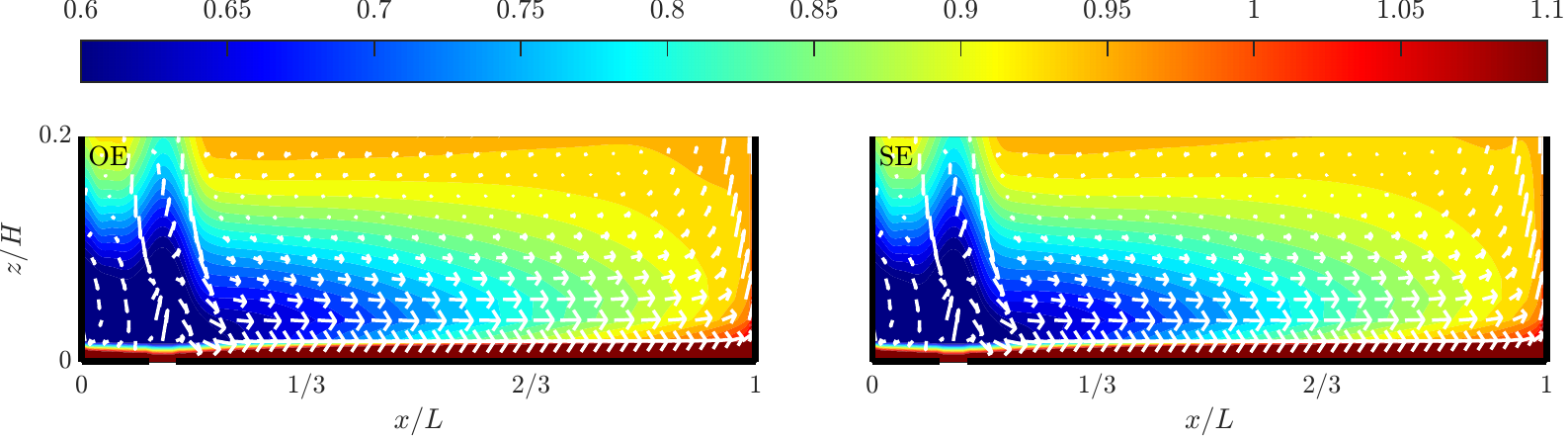}
         \label{fig:CFD_BL1}
     \end{subfigure}
        \caption{Scaled buoyancy fields $\langle \overline{\Delta \mathcal{T}}\rangle_y$ from the numerical simulations presented only in the lower portion of the room, $0 \leq z/H \leq 0.2$. Velocity vectors are overlaid in white. {The opposite-ended (OE) configuration is shown in the left column and the single-ended (SE) configuration in the right column.} Note that in each figure thick black lines mark the confined edges of the room, gaps {mark} the positioning of the low-level inlet vents.}
        \label{fig:CFD_buoyancyfieldBL}
\end{figure}

{Before examining the details of the flows established, we note the following similarities to the flows typically described as horizontal convection \citep{Hughes08}. Firstly, the overall sense (direction) of the circulation in the room is set by the location of the inlet relative to the distributed heat/buoyancy forcing --- i.e. the sense of the circulation of the large-scale convective flow is set by the largest horizontal buoyancy contrast. Secondly, in a steady state, the strength of this circulation adjusts such that the incoming ambient air acquires sufficient buoyancy to rise against the far endwall, thus coupling the bulk buoyancy with the large-scale convective flow in the room. Whilst the coupling between the bulk buoyancy and the large-scale convective flow remains, the orientation of the inlet (e.g lying horizontal in the floor or vertical in a wall) can be expected to result in only minor changes in the flow field within the room. Furthermore, we expect this coupling to remain until the momentum flux of the inlet is sufficiently forced; which it is not in any of our cases, since the flow is naturally forced by the buoyancy. Thirdly, as a consequence of the fact that the incoming ambient air acquires sufficient buoyancy to rise against the far endwall, and hence the bulk buoyancy is coupled with the large-scale convective flow within the room, this convective flow can be anticipated to run along the full length of the floor and full height of the room, regardless of the room aspect ratio. Hence the similarities observed between all the flows we examined can be expected. Moreover, the coupling of the large-scale convective flow which rises up the far wall, with the buoyancy that drives the ventilation, also underlies many of the differences we observed. For example, in the opposite-ended case the outlet vent is relatively close to the location where the large-scale convective flow rises; for the single-ended case the large-scale convective flow must return back along the length of the ceiling before reaching the outlet vent --- the latter results in a large-scale overturning motion that fills the room, typical of classical horizontal convention flows.}

\begin{figure}
    \centering
    \includegraphics[width = \textwidth]{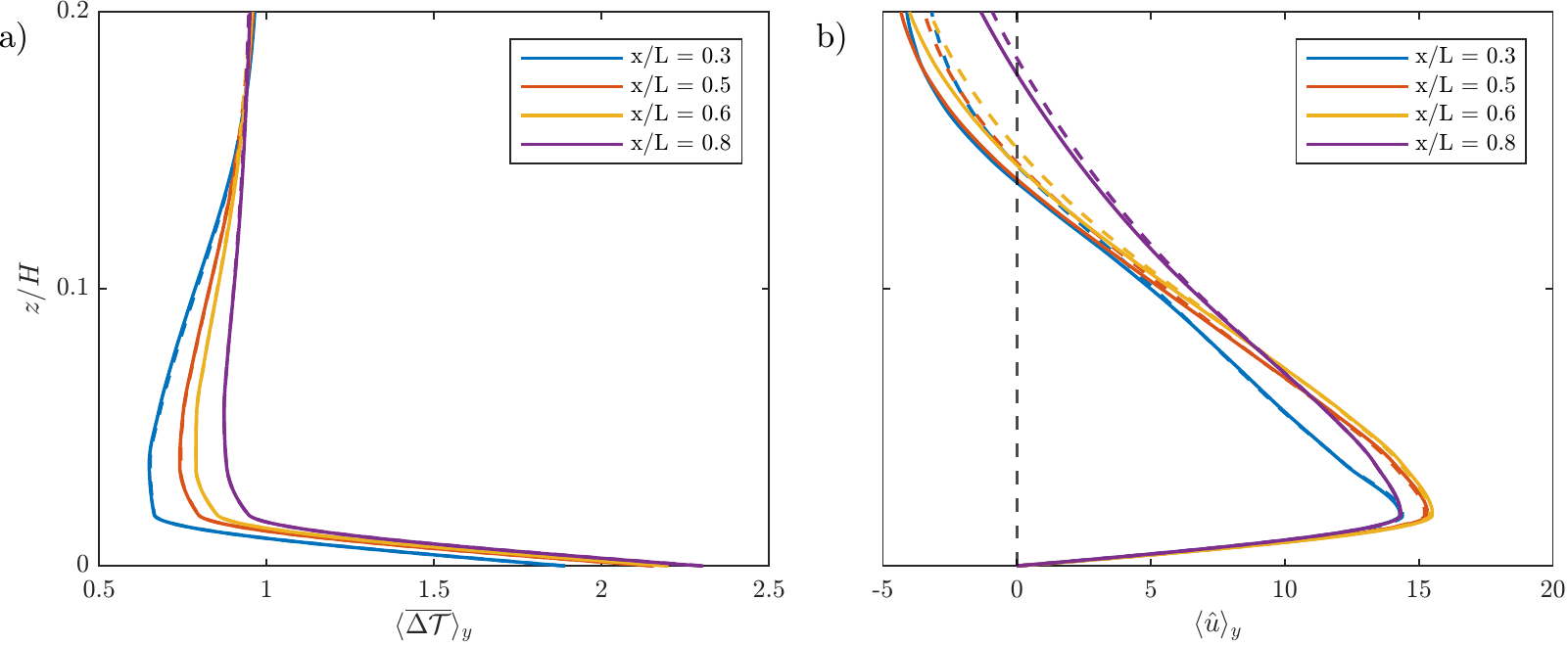}
    \caption{Vertical profiles within the lower portion of the room ($0 \leq z/H \leq 0.2$) from both the opposite-ended configuration (solid lines) and the single-sided configuration (dashed lines): a) width-averaged scaled buoyancy $\langle\overline{\Delta \mathcal{T}}\rangle_y$ and b) scaled horizontal velocity $\langle \hat{u} \rangle_y$. In both {panes} profiles are shown at four horizontal locations: $x/L = \{0.3, 0.5,0.6, 0.8\}$ {along the length of the room.}}
    \label{fig:BL_profiles}
\end{figure}

To begin examining the physics of the flows examined herein, we present the scaled buoyancy fields in the lower portion of the room $0 \leq z/H \leq 0.2$ (figure \ref{fig:CFD_buoyancyfieldBL}). {Regardless of the locations of the outlet vents} the similarities between the two configurations are striking. To elucidate the buoyancy driven flow that arises, we overlay white arrows representing the velocity vectors of the width-averaged flow field. The magnitudes of, and flow patterns indicated by, these velocity vectors are very similar. In both cases, the cold air {drawn} in through the vent in the floor creates a rising fountain of cooler air {near the inlet with relatively large vertical velocity}, fluid from which slumps back and propagates along the floor in a cool gravity current. Fluid in this current is continually warmed as it propagates by the heat emitted from the floor. In a manner similar to evolution of the convective mixed layer observed in horizontal convection \citep{Mullarney04}, this {heat} is mixed into, rather than penetrates through, the gravity current. In the region close to the endwall, the fluid within the current has warmed to a temperature of $\langle \overline{\Delta \mathcal{T}}\rangle_y \approx 1$, i.e. close to that of the well mixed state. Figure \ref{fig:BL_profiles}a) illustrates the horizontal evolution of the temperature within the gravity current by presenting vertical profiles of the scaled buoyancy at four different horizontal locations.

The evolution of the (scaled) velocity profile, $\hat{u}(z)$, within this gravity current is further shown in figure \ref{fig:BL_profiles}b). We present the ({width}-averaged) horizontal velocities, $\langle \hat{u} \rangle_{y}$, scaled by the average velocity in an idealised two-layer horizontal flow within the room, i.e. $U = 2Q/(W H)$. One can see from the height at which the horizontal velocities change sign, that the gravity current occupies only around the bottom 15\% of the room. Moreover, the magnitudes of the scaled velocities integrated over this bottom region indicate that the full ventilation flow $Q$ is accommodated within the gravity current. {We} note that if one simply looks at peak horizontal velocities, the gravity current accelerates slightly from $x/L=0.3$ towards the middle of the centre ($x/L=0.5$ and $x/L=0.6$) before then decelerating slightly towards $x/L=0.8$.

\begin{figure}
\centering
     \begin{subfigure}[h]{\textwidth}
         \centering
         \includegraphics[width=\textwidth]{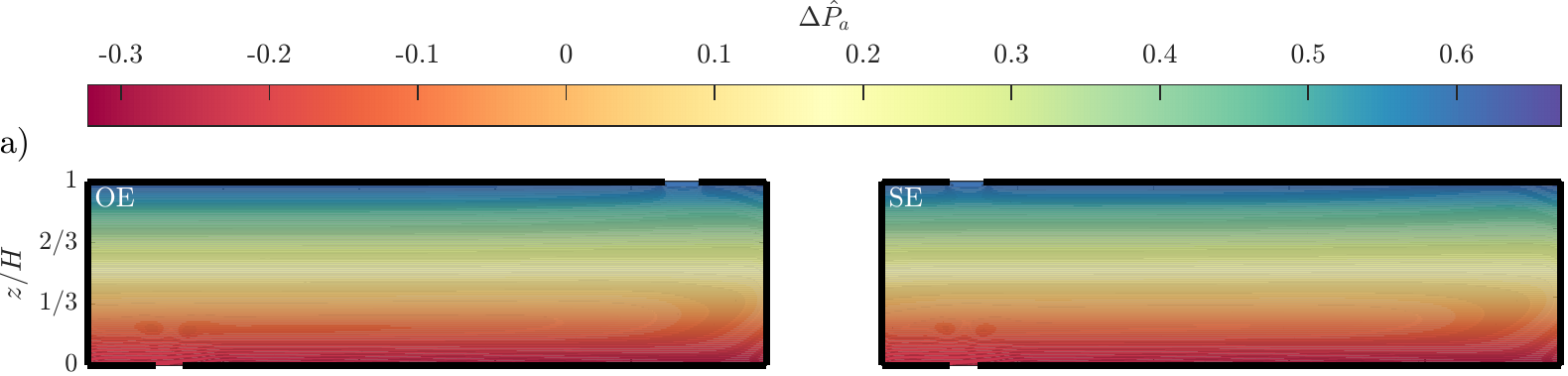}
     \end{subfigure}
    \vspace{-0.2pt}     
     \begin{subfigure}[h]{\textwidth}
         \centering
         \includegraphics[width=\textwidth]{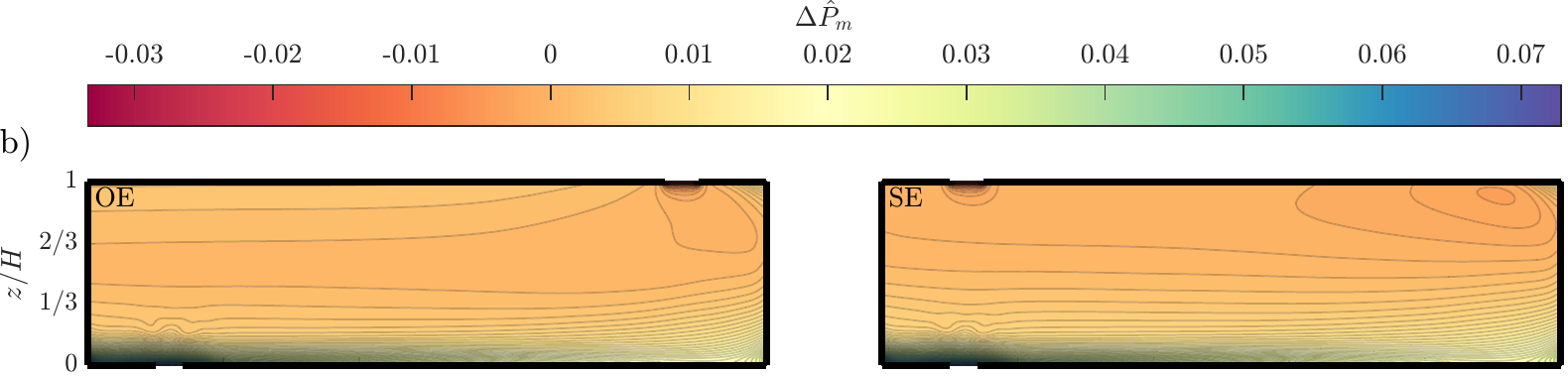}
     \end{subfigure}
       
         \begin{subfigure}[h]{\textwidth}
         \vspace{10pt}
         \centering
         \includegraphics[width=\textwidth]{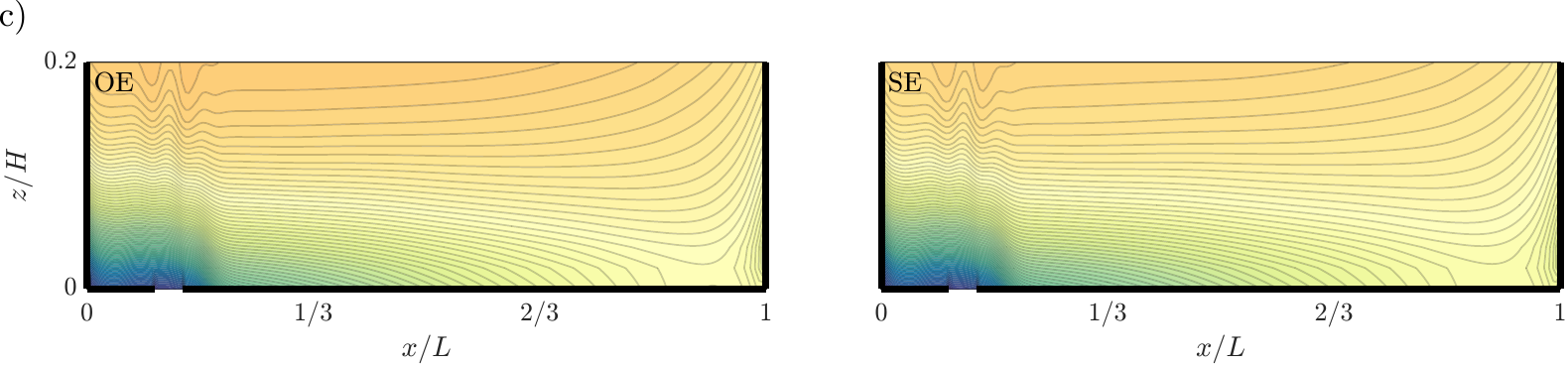}
         \label{fig:CFD_pressureBL}
     \end{subfigure}
        \caption{{Width}-averaged pressure fields within the room scaled by $\Delta P = \rho_a \left(F \, H/A^* \right)^{2/3}$. Top row, a) scaled pressure difference relative to the ambient pressure $\Delta \hat{P}_a = [P(x,z)-P_a(z)]/\Delta P$. Middle row, b) scaled pressure anomaly relatively to the pressure within a well-mixed room $\Delta \hat{P}_m = [P(x,z)-P_m(z)]/\Delta P$, where $P_m(z) = P_a(z) - \rho_a \, z \, F / Q$, and bottom row, c) scaled pressure anomaly $\Delta \hat{P}_m$ in the lower portion of the room, $0 \leq z/H \leq 0.2$.
        The left-hand {panes} show the opposite-ended {(OE) }configuration, the right-hand {panes show} the single-ended {(SE)}. Thick black lines mark the confined edges of the room, gaps {marks} the positioning of the low- and high-level, inlet and outlet vents, respectively.
        }
        \label{fig:CFD_pressure} \label{fig:press_a}\label{fig:press_m} \label{fig:press_mb}
\end{figure}

To understand how flows in the two different room configurations {can} be so similar in {the lower} portion of the room and yet {be} quite different {higher in the room}, we investigate the pressure field driving the flow.
Figure \ref{fig:press_a}a) shows the {width}-averaged pressure within the room $\langle \overline{P} \rangle_{y} = P(x,z)$ relative to the ambient pressure outside, $P_a(z)$. We present all pressures scaled by the natural pressure scale $\Delta P = \rho_a \left(F \, H/A^* \right)^{2/3}= \rho_a \, b_m \, H$, where $b_m$ is the buoyancy attained in the well-mixed state{, so} that $\Delta P$ represents the full scale of the available driving pressure from the buoyancy source within {the} room. Results for $\hat{\Delta P_a} = (\langle \overline{P}\rangle_y - P_a(z))/\Delta P$ (figure \ref{fig:press_a}a) {show that} the pressure differences {are} broadly horizontally homogeneous, span a range of magnitude equal to unity, and the neutral pressure level $z_{npl}$ \citep[at which $\hat{\Delta P_a}=0$, see e.g.][]{Connick20} lies {below} the middle of the room, as expected when the greatest restriction to flow is provided by the high-level vent, note $A_l/A_h=0.5$, see \S\ref{sec:num}. {This form of the} pressure field indicates why the bulk flow rates can be reasonably predicted by assuming a well-mixed state {since this is driven by the integrated buoyancy}. 

To understand the internal flows within the room, it proves useful to examine the width-averaged pressure relative to the pressure that would be present if the room were truly well-mixed, which we denote $P_m(z)$. To do so, we exploit knowledge of the neutral pressure level which we attain from horizontally averaging $\hat{\Delta P_a}$ (figure \ref{fig:press_a}a), giving $z_{npl}/H \approx 0.3$ for both the opposite-ended and the single-ended {cases}. We define the pressure in the well-mixed state to be $P_m(z) = P_a(z) + \rho_a \, (z-z_{npl}) \, F / Q$, so that the scaled pressure anomalies are
\begin{equation}
     \hat{\Delta P_m} = \frac{\langle \overline{P}\rangle_y - P_m(z)}{\Delta P}  \, ,
\end{equation}
and at the height $z=z_{npl}$, $\hat{\Delta P_m}=0$. Figure \ref{fig:press_m}b) shows this scaled pressure anomaly and {shows} that throughout the bulk of the room $\hat{\Delta P_m} \approx 0$ with the strongest deviations occurring close to the lower inlet vent. It is notable that this data show there to be an adverse horizontal pressure gradient within the upper portion of the room, such that any fluid rising at the far wall which overshoots the vent, might be expected to remain within the region above the lower inlet vent. Focusing on the lower portion of the room figure \ref{fig:press_mb}c) presents this data for the range $0 \leq z/H \leq 0.2$. {While f}igure \ref{fig:press_a}a) showed that, on average the scaled vertical pressure gradients remain close to unity in this bottom region, the data presented in figure \ref{fig:press_mb}c) show that the vertical gradients in the pressure anomalies from the well-mixed state are relatively strong in this region. This is especially true in the region above the vent where $\hat{\Delta P_m}$ changes by approximately -0.07 over the height $0 \leq z/H \leq 0.2$, giving a vertical gradient of $\ud \, \hat{\Delta P_m} / \ud (z/H) \approx -0.3$ for the pressure anomalies from the well-mixed state, \textit{cf.} $\ud \, \hat{\Delta P_a} / \ud (z/H) \approx 1$ for the pressure differences relative to ambient. However, within the bottom region the flow is driven by horizontal pressure gradients which are far smaller, $\ud \, \hat{\Delta P_m} / \ud (x/H) \approx -0.02$. Despite these horizontal pressure gradients being nearly two orders of magnitude smaller than the vertical pressure gradients that drive the bulk flow, these horizontal pressure gradients are sufficient to establish the large-scale horizontal convection within the rooms which have significant implications for the occupant experience within the room. 

\begin{figure}
     \centering
     \begin{subfigure}[b]{\textwidth}
         \centering
         \includegraphics[width=\textwidth]{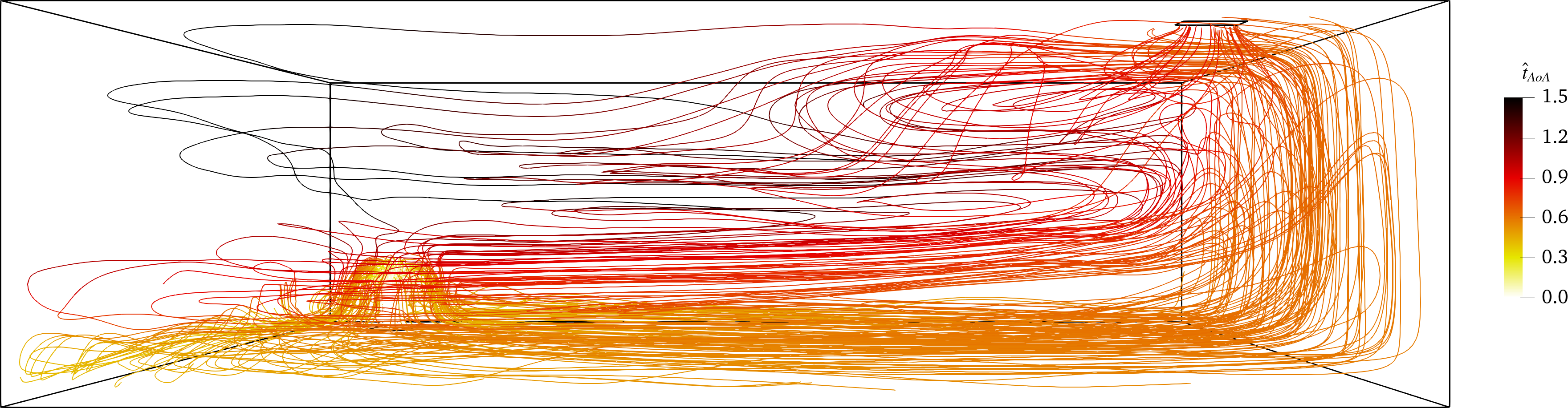}
         \caption{Opposite-ended}
     \end{subfigure}
     ~
     \begin{subfigure}[b]{\textwidth}
         \centering
         \includegraphics[width=\textwidth]{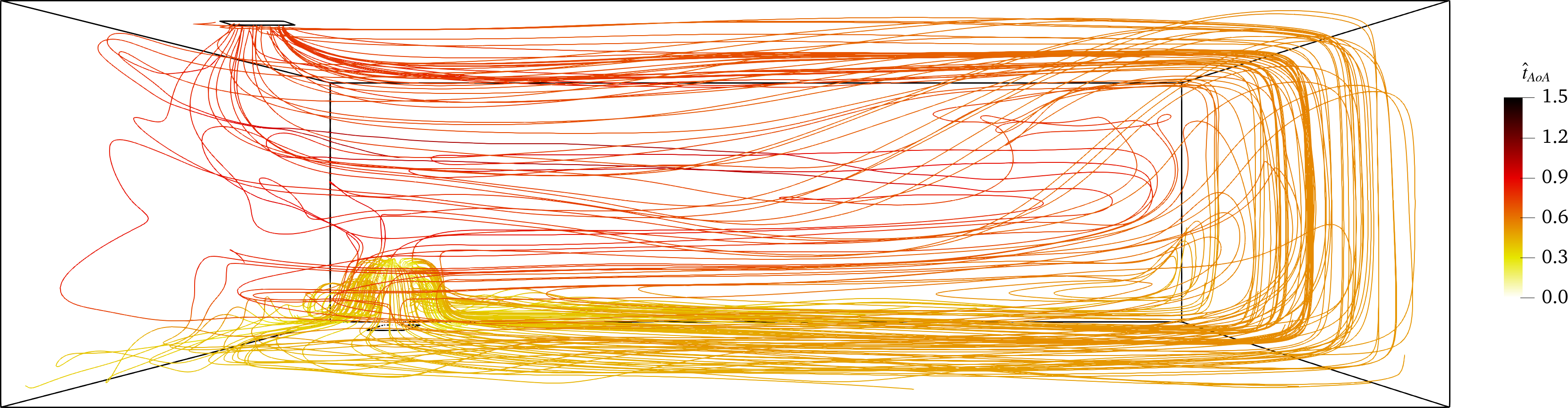}
         \caption{Single-ended}
     \end{subfigure}
     \caption{Three dimensional illustration of the streamlines in a) the opposite-ended configuration, and b) the single-ended configuration. In both cases, the room is viewed from the front, and 100 streamlines which originate from (i.e. seeded at) the low-level inlet vent {are shown.} The streamlines are coloured by the scaled age of air. {Note that, as with the other figures included herein, data is only shown from within the room, see figure \ref{fig:Mesh} for an illustration of the room within the full computational mesh.}}
     \label{fig:CFD_streamline_front}
\end{figure}

{Although }the flow patterns and temperatures in the bottom portion of the room are very similar between the two configurations, the streamlines (figure \ref{fig:CFD_streamline_front}) {show significant} differences in the flow fields between the two configurations. In the case of the single-ended configuration, the relatively cool air rises at the inlet but falls back due to its buoyancy and is then carried over the full floor area, being warmed as it travels {towards the far endwall.} {After rising} near the right-hand {end}-wall, {the air} (broadly speaking) travels back {along the length of the room} in the upper {part} of the room and out through the high-level vent. {In c}ontrast, {in} the opposite-ended configuration, the warmed fluid rises in the region near the right-hand {end}-wall {and} predominantly travels up and out the high-level vent at the far {end} of the room. As a consequence, a relatively low volume flux of fluid travels through the region in the upper portion of the room above the region of the inlet vent, {creating} a re-circulation or `dead' zone. We now consider these findings in the context of the potential impact on occupants. {Examination of the vorticity field highlighted similarities between the opposite-ended and single-ended configurations within the lower layer; similarities which were reflected, when viewed relative to the outlet, in the upper portion of the room too.}

\subsection{Implications for thermal comfort and exposure} \label{sec:impli}

\begin{figure}
    \centering
    \includegraphics[width=\textwidth]{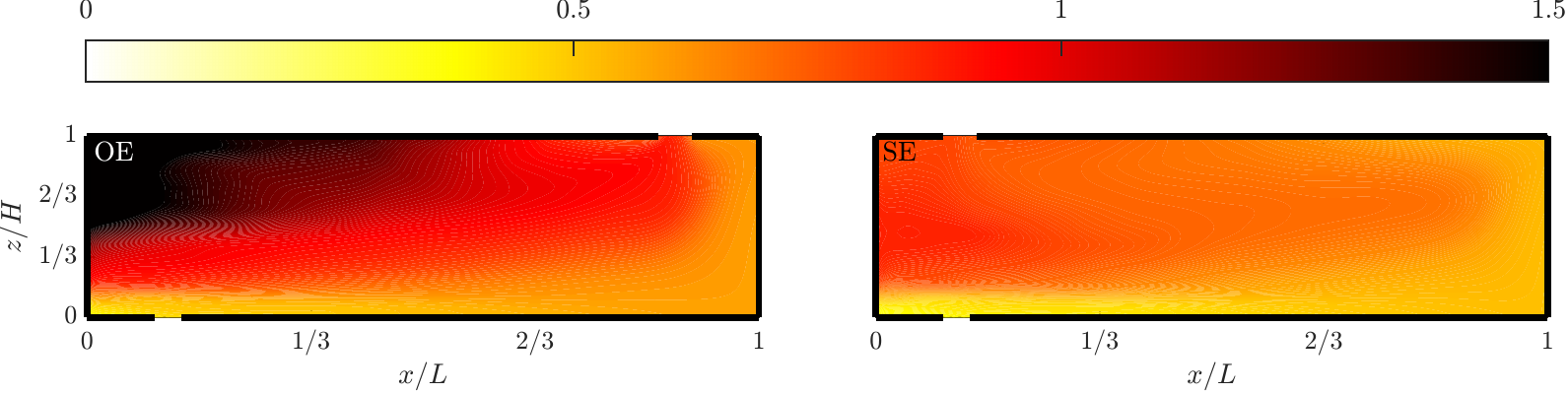}
    \caption{The (scaled) age of air $\hat{t}_{AoA}$ at a plane through the outlet{s} ($y = 0.8 W$); within the opposite-ended (left {OE}) and single-ended (right {SE}) configurations.}
    \label{fig:CFD_AoAOutletPlane}
\end{figure}

{Although} the buoyancy field indicates variation in occupant thermal comfort, a more pertinent {measure of} indoor air quality and occupant exposure to pollutants is the scaled age of air {$\hat{t}_{AoA}$} which determines how long fluid has remained within the room relative to the bulk ventilation air change rate $Q/V$ \citep[see \S \ref{sec:methods} and][]{Sandberg81}. Figure~\ref{fig:CFD_AoAOutletPlane} plots the scale age of air on a vertical slice through the central plane of the outlet vent, data is presented for both configurations --- differences are notable. In the opposite-ended configuration, the age of air in the upper part of the room above the lower inlet vent is $\approx 1.5$, clear evidence of the impact of the dead-zone {shown in the top left corner of figure \ref{fig:CFD_streamline_front}a)}, with no such counterpart in the single-ended configuration. Despite these differences we find that the age of air at the outlet is almost identical for both configurations, {having values } 0.775 {and 0.774} for the opposite-ended and the single-ended {cases, respectively, (see} figure \ref{fig:CFD_AoAOutletPlane}). {It is unclear whether this agreement is a coincidence or a feature of the flows having the same ventilation rate.}  

\begin{figure}
\centering
         \centering
         \includegraphics[width=\textwidth]{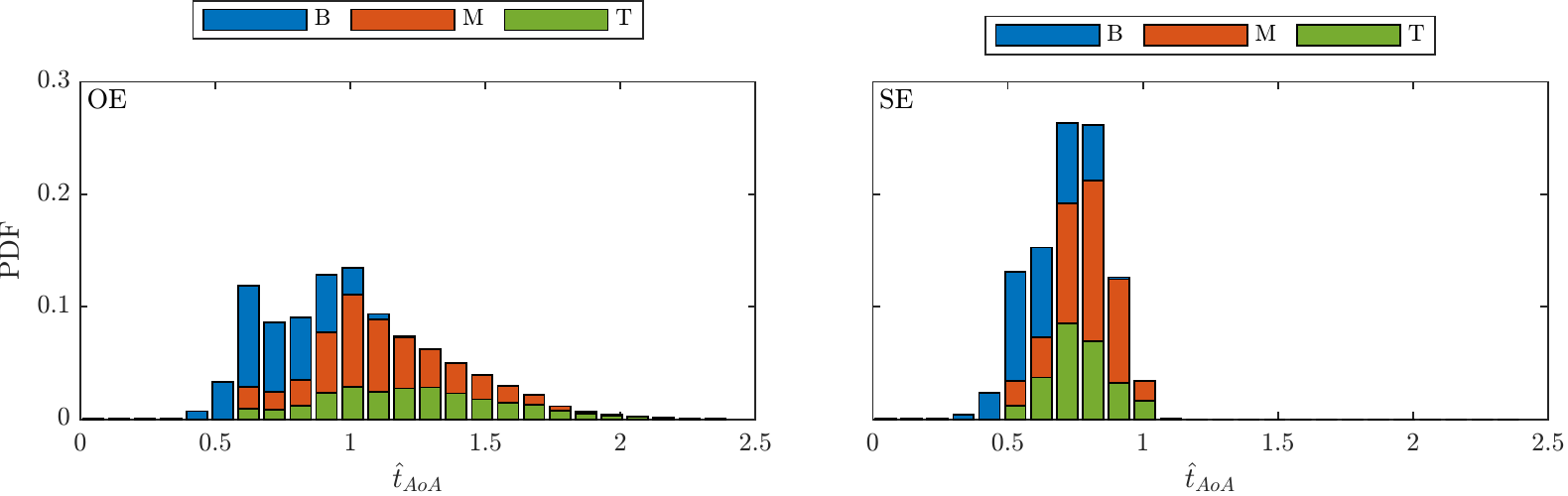}
         \caption{Histogram of the { scaled} age of air {$\hat{t}_{AOA}$} within the room. The colours highlight the portion associated with three vertical regions: {green - }top, T($2H/3\leq z \leq H$); {red - }middle, M($H/3\leq z \leq 2H/3$); and {blue - }bottom, B($0\leq z \leq H/3$). {In the opposite-ended (OE) configuration the mean in each vertical region is $\mu_T = 1.24$, $\mu_M = 1.11$ and $\mu_B = 0.74$ with respective standard deviation $\sigma_T = 0.34$, $\sigma_M = 0.26$ and $\sigma_B = 0.16$. In the single-ended (SE) configuration it is $\mu_T = 0.77$, $\mu_M = 0.79$ and $\mu_B = 0.63$ with $\sigma_T = 0.11$, $\sigma_M = 0.11$ and $\sigma_B = 0.12$.}}
         \label{fig:CFD_PdfAoA}
\end{figure}

Histograms of which are plotted in figure \ref{fig:CFD_PdfAoA}. The mean (scaled) age of air within the room {is} $1.01 \pm 0.35$ for the opposite-ended configuration and $0.71 \pm 0.14$ for the single-ended (with the bound indicating one standard deviation) -- these differences in mean and variance are both statistically significant. Figure \ref{fig:CFD_PdfAoA} also illustrates that, for the single-ended case, only about 6\% of the air within the room exhibits $\hat{t}_{AoA} \gtrsim 1$; {i. e.} the fluid within the room that has had time to fully circulate the room once. All other fluid within the room must be, on average, in the process of being efficiently circulated around and out of the room. In contrast, not only does the age of air distribution for the opposite-ended case show a much wider variance, {with }around 20\% of the air within the room satisfying $\hat{t}_{AoA} \gtrsim 1.5$. This suggests the potential for `dead-zones' (regions in which fluid {is }largely recirculated) {is} more prevalent in the opposite-ended configuration than in the single-ended configuration. Note also that all of the fluid with $\hat{t}_{AoA} \gtrsim 1$ lies in the upper two-thirds of the room but, perhaps more pertinent for exposures, much of this more stale air lies in the middle two-thirds of the room, $0.9\textrm{m} \leq z \leq 1.8\textrm{m}$, which includes the breathing zone.

\section{Conclusions} \label{sec:conc}

{This paper reports the results of laboratory experiments and corresponding computations of the flow produced by a uniformly heated floor in a ventilated room. The room is connected to the exterior by upper and lower vents, and the heated floor produces displacement ventilation where cool air enters through the low-level vent and leaves via the high-level vent. 
We focused on the effects of the relative location of the upper vent in relation to the lower vent by considering two cases: the single-ended case where the outlet is at the same end of the room as the inlet, and the opposite-ended case where the outlet vent is at the other end of the room. Two rooms were studied, a relatively short room with length 1.7 times the height and a long room with length 3.7 times the height. 

{In all cases, large-scale convective flows travelled along the full length of floor and up the far wall, and all our findings were, within the bounds tested, insensitive to whether the momentum of the incoming {cool air} is vertical or horizontal, and insensitive to the aspect ratio of the room --- these facts can be understood as follows.
Firstly, we observe that the overall sense of the circulation in the room is set by the inlet location relative to the distributed heat/buoyancy forcing (due to this setting the vertical level of the greatest horizontal buoyancy contrast). The steady state strength of circulation adjusts such that the incoming ambient air acquires sufficient buoyancy to rise against the endwall, thus coupling the bulk buoyancy with the large-scale convective flow in the room; a coupling which, for these buoyancy-driven flows, results in the momentum flux associated with the inflow being of little consequence irrespective of the orientation of the vent. Furthermore, since the incoming ambient air acquires sufficient buoyancy to rise against the endwall, this large-scale convective flow can be expected to run along the full length and height of the room, regardless of the room aspect ratio.

A key conclusion of this research, which follows from the physics of this large-scale convective flow,} is that even though, as expected the ventilation rate of the room does not depend on the location of the outlet vents, the flow within the upper part of the room is quite different in the single-ended and opposite-ended cases. {In the opposite-ended case, the outlet vent is relatively close to the location where the large-scale convective flow rises; for the the single-ended case the large-scale convective flow must return along the length of the ceiling before reaching the outlet vent --- the latter results in a large-scale overturning motion that fills the room, typical of classical horizontal convection flows.} In the single-ended case the upper regions of the room are relatively uniform in temperature, while in the opposite-ended case there are significant differences in temperature from one end to the other. In this latter case the }results exhibit spatial inhomogeneities of the buoyancy scalar {of the order of} $20\%$ of the mean buoyancy. 

These inhomogenities have implications for the exposures that occupants might experience if positioned at different locations within the room. These findings are {supported} by {calculations of} the age of air and the streamline patterns, which {show} that {a} relatively stagnant `dead-zone' {is} formed above the inlet in the opposite-ended case. While most of the air rising up the far endwall exits at the outlet vent a small amount travels back towards the inlet in the upper two-thirds of the room and accumulates there. Over time this leads to an accumulation of buoyancy in this upper region. The horizontal variation would imply that pollutant entering from outside would also accumulate and have concentrations above the mean for the room as a whole. Similarly, these variations have implications for indoor sources of pollution and the spread of airborne biological material released inside the room.

Although the cases considered in this paper are somewhat idealised, one should expect the pattern of the large-scale convection observed to be relatively robust and hence one could expect flow patterns similar to those reported herein to be realised in some operational rooms. For example, based on the understanding of horizontal convection \citep[e.g.][]{Hughes08}, the flows observed are expected to be robust to the inclusion of additional localised heat sources. This can be expected because: firstly, the timescale for the fluid above the horizontal buoyancy source to be swept to the end wall of the room is comparable with the timescale of rise through the lower stratified region of buoyant parcels so that, with the associated vertical and horizontal length scales, the momentum of the horizontal flow remains dominant in organising the overall circulation; and secondly, the lower stratified region acts to confine vertically parcels that are initially buoyant at the horizontal boundary; and finally, in the presence of such statistically steady convective flows, the buoyancy of the warmest fluid in the room interior is approximately equal to (i.e. is set by) that of the fluid above the bottom horizontal boundary adjacent to the far endwall. These effects will act to render the convective circulation relatively robust to perturbations that arise due to the presence of localised heat sources. As such, our findings have three further implications that are worth noting. First, point measurements of {temperature, carbon dioxide and pollutant }concentration, commonplace in the assessment of building spaces, {should be treated with caution as they may not be representative of the room as a whole. Second, the results show} in the cases investigated, that a `single-ended' displacement design of positioning all low- and high-level vents at one end of the room provides a more {uniformly} mixed indoor environment. This might deserve consideration when revising any design guidance which typically promotes opposite-ended strategies for natural ventilation, as a result of expected contributions by the wind enhancing the bulk ventilation flow rate \citep[e.g.][]{CIBSEAM10}. {Third, w}e acknowledge that our findings increase the complexity of making useful predictions {from simple design rules}. However, they do highlight the rich variety of dynamics that can arise in indoor air flows and the value of deepening our understanding as we spend ever more time indoors. Ultimately, our results set a challenge to understand and classify indoor airflows into a broader set of classes and, for each, to quantify the typical magnitudes of the variations in scalar concentrations that can be expected to arise.

\appendix

{
\section{Details of the laboratory experiments} \label{app:exp}

{A cuboid}, as a small scale model of a room, was connected to the ambient environment {(a visualisation tank of cross-section $\SI{1.3}{m}\times\SI{1.3}{m}$ filled with fresh water to a depth of $\SI{1.25}{m}$)} by two openings, a low-level inlet vent (aligned in the vertical plane much like a small doorway) and a high-level outlet vent, 
the position of the vents are illustrated in figure \ref{fig:exp_t_avg_buoyancy_fields} -- see table \ref{tab:params} for details.
The high-level outlet was positioned in the ceiling approximately two thirds of the way along the room from the inlet, and hence the bulk flow must be both upwards and across the room.
{A saline solution of buoyancy $b_{0} = \SI[parse-numbers=false]{0.35\pm0.01}{\m\per\square\s}$ (the density of which was measured using an Anton Paar DMA 4500 density meter with precision $1 \times 10^{-5}$\,g/cm$^3$) was supplied to the scale model at floor level across a porous plastic (HDPE) panel (Vyon F $\SI{6}{mm}$ thickness) with a flow rate of $Q_{0}=\SI[parse-numbers=false]{1.67\times10^{-5}}{\m\cubed\per\s}$ (controlled by a peristaltic pump [Cole Parmer Masterflex L/S] and monitored using an in-line ultrasonic flow meter [Atrato 740-V10-D]).
The source volume flux was always less than 1\% of the predicted bulk ventilation rate \citep{Gladstone00} and was therefore assumed to be largely representative of the flow pattern in the full-scale application.
A two-chamber system was used to provide a source buoyancy flux that was evenly distributed over the floor area. The principle of this system was to ensure that any horizontal pressure gradients across the source where negligible compared to the pressure gradients imposed by the pump to drive the flow across the porous plastic panel of the source \citep[see][for further details]{Higton2022}.

The source solution was dyed with a red food colouring, Allura red (AC, E129), to enable t}he width-averaged two-dimensional buoyancy field $\langle b \rangle_y=b(x,z,t)$ to be measured using the dye-attenuation technique \citep{Cenedese97,Allgayer12}.
{The model `room' was illuminated from behind using a $\SI{1}{m}\times\SI{1}{m}$ light panel (Applelec LED light sheet), and the light attenuated by the dyed fluid in the model was measured by a Sony A7Rii digital camera (14-bit depth RGB images with a spatial resolution of approximately $\SI{0.14}{\mm\per pixel}$).
}
Note that, experimental data directly adjacent to the walls are compromised due to parallax errors, camera viewing restrictions, and reflections from the walls and hence these regions are excluded from the results presented -- these regions are highlighted in figure \ref{fig:exp_t_avg_buoyancy_fields}. 

\section{Details of the numerical simulations} \label{app:sim}
Numerical simulations were conducted with OpenFOAMv2106 using the transient \texttt{buoyant\-Pimple\-Foam} solver and the $k-\omega$ SST turbulence model. The chosen OpenFOAM version also incorporates the effects of buoyancy on turbulence production by using the \texttt{buoyancy\-Turb\-Source} finite volume option \citep{OFbuoyancy}. Full details of the governing equations and methods of solution are available in the OpenFOAM documentation \citep{OFv2106}.

The computational domain includes the room, a cube of dimensions 10\,m $\times$ 5.5\,m $\times 2.7$\,m, along with two exterior boxes linked to the room through inlets and outlets (see figure \ref{fig:Mesh}). These two boxes were included in order to properly model the effect of flow at the two openings and the resulting flow in the room. The bottom exterior box is centred around the inlet and has dimensions 2.8\,m $\times$ 2.5\,m $\times$ 1\,m. The top exterior covers the same surface as the room with dimensions 10\,m $\times$ 5.5\,m $\times$ 3.7\,m. The inlet and outlet connecting the exterior boxes to the room have the same size as the vents with respective areas $A_l$ and $A_h$, and height 0.2\,m. The sensitivity of the simulations to the size of the external boxes was tested and these dimensions were chosen as they did not impact the results. The mesh is defined as a perfect orthogonal mesh with $\Delta x=\Delta y =\Delta z =0.05 $\,m in the room, inlet and outlet and $\Delta x=\Delta y =\Delta z =0.1 $\,m in the exterior, requiring computations to be executed on 1.4 $\times 10^6$ hexahedral grid cells. A grid convergence study was also performed: the mesh used in this study accurately represents the bulk flow as well as the buoyancy and tracer distribution when compared to a finer mesh (with $\Delta x=\Delta y =\Delta z =0.025 $\,m), it allows captures the variations in the flow pattern observed in the single-ended and opposite-ended configurations. Full details are presented by \citet{Vouriot2022}.

\begin{figure}
    \centering
    \includegraphics[width=\textwidth]{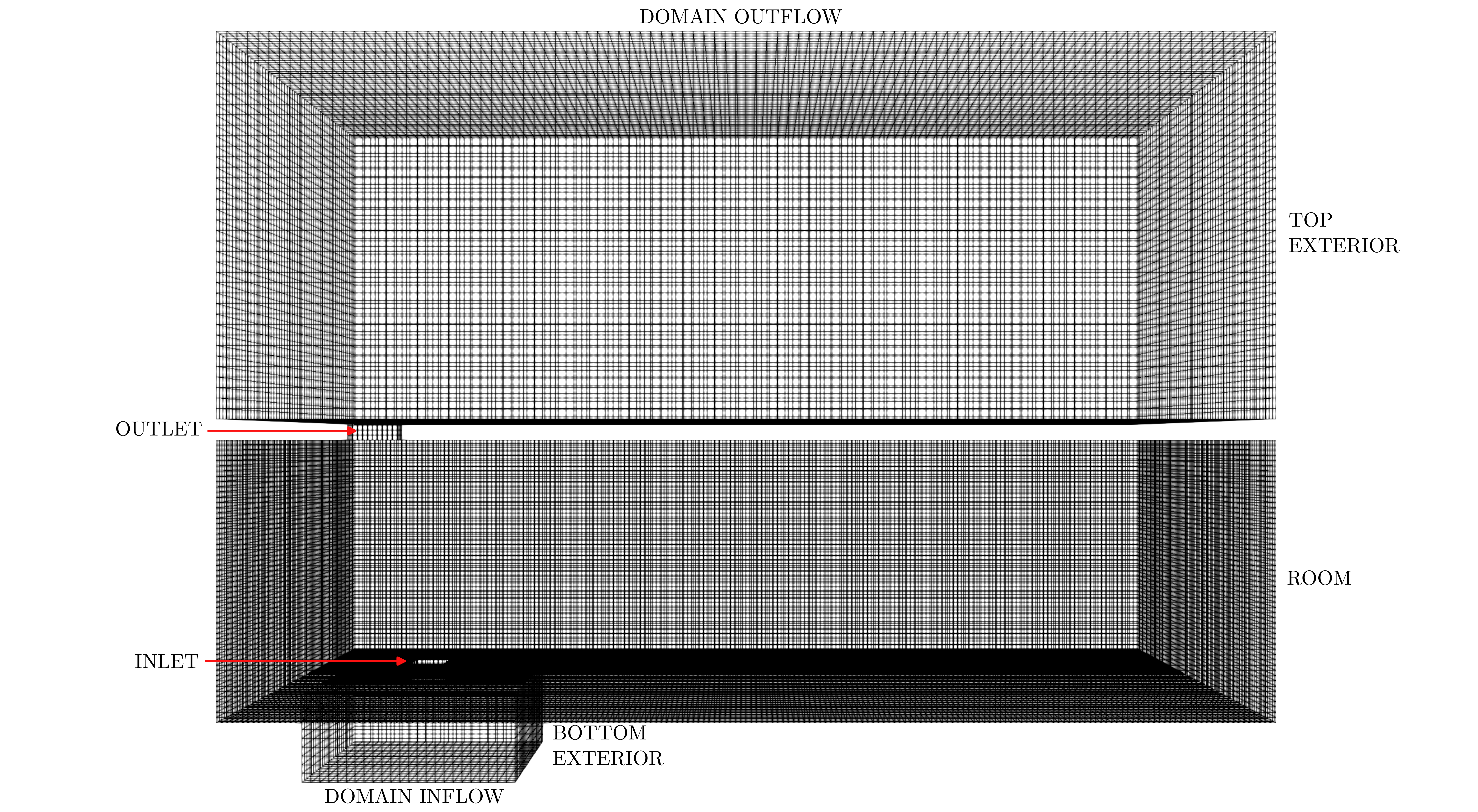}
    \caption{Computational grid used in the numerical simulations shown for the single-ended configuration, including the modelled exterior, inlet, outlet and room.}
    \label{fig:Mesh}
\end{figure}

At all walls, including the room and the exterior boxes' (apart from the domain inflow and outflow), a no-slip velocity boundary condition is used. All walls are also assumed to be adiabatic, with the exception of the floor where a constant 6,200\,W heat input is imposed. At the domain inflow the temperature is set to ambient (defined as 278K in this simulation) and at the domain outflow a Neumann boundary condition is used setting the gradient of the temperature. Velocity boundary conditions at the inflow and outflow are calculated from the pressure field where the pressure difference $\Delta p_0 = -\rho_a g H_{domain}$ is imposed across the domain inflow and outflow. The boundary layer region is not fully resolved, instead wall functions are used with standard parameters for: the turbulent thermal diffusivity, the turbulent kinetic energy, the turbulent viscosity and specific dissipation rate. The stratification is first established using a steady run, the simulations are then run with a transient solver for 8,600\,s with statistics averaged only over the last 3,600\,s. A transient solver had to be used due the presence of long-time period fluctuations (potentially caused by internal wave modes) observed in the results generated by steady solvers. Second-order backward finite difference time-marching and second-order central differencing schemes are used. Full details and the specific OpenFOAM set-up used are described by \citet{Vouriot2022}.

\section{Sensitivity of the dimensionless volume flux on the choice of discharge coefficient} \label{app:sen}

For completeness, the values of $A^*$, and thereby $\hat{Q}$, depend on $C_d$, and in table \ref{tab:params} we took $C_d = 0.65$. Should we have taken $C_d=0.6$ then $1.00 \leq \hat{Q} \leq 1.08$ for the experiments and $1.05 \leq \hat{Q} \leq 1.12$ for the simulations. Conversely, taking $C_d=0.7$ would then give $0.91 \leq \hat{Q} \leq 0.98$ for the experiments and $0.95 \leq \hat{Q} \leq 1.02$ for the simulations. 
}

\clearpage

\paragraph{Acknowledgements}
CVMV, HCB, and MvR gratefully acknowledge Fred Mendonça and his team at OpenCFD for providing support and advice on setting up the numerical simulations. CVMV, HCB, and PFL acknowledge the support of the SAMHE Project consortium.

\paragraph{Funding Statement}
The PhD work of CVMV and TDH, data from which inspired and underpinned this work, were funded by the Engineering and Physical Sciences Research Council (EPSRC) via the Imperial College London Centres for Doctoral Training in: Fluid Dynamics Across Scales (grant EP/L016230/1), and Sustainable Civil Engineering (grant EP/R512540/1), respectively. The contributions from HCB and PFL were funded by the EPSRC COvid-19 Transmission Risk Assessment Case Studies - education Establishments (CO-TRACE) project, and the extension, School Air quality Monitoring for Health and Education, SAMHE (both under grant EP/W001411/1). HCB was further supported in his role within WP2.2.2: Ventilation Effects, within the PROTECT COVID-19 National Core Study on transmission and environment, managed by the Health and Safety Executive on behalf of HM Government. Computational resources were provided by the UK Turbulence Consortium (grant EP/R029326/1).

\paragraph{Declaration of Interests}
The authors declare no conflict of interest.

\paragraph{Author Contributions}
Through their PhD research, CVMV performed all simulations, and TDH performed all experiments --- including carrying out their respective analysis through to the creation of the plots included. HCB supervised the PhD work of CVV and TDH, with co-supervision provided by MvR and GOH, respectively. HCB supervised the data presentation and wrote the manuscript. PFL identified the value of the finding reported herein, thus inspiring the creation of this paper. All authors edited the text, providing invaluable insights and suggestions for the analysis.

\paragraph{Data Availability Statement}
Raw data are available from the corresponding author
(HCB).

\paragraph{Ethical Standards}
The research meets all ethical guidelines, including
adherence to the legal requirements of the study country.

\paragraph{Supplementary Material}
Supplementary information are available
at \url{https://doi.org/10.1017/flo.2021.XXX}.

\bibliographystyle{jfmF} 
\bibliography{bibl_USE.bib}

\end{document}